\magnification = \magstep1
\baselineskip = 24 true pt
\hsize = 16 true cm
\vsize = 22 true cm
\centerline {\bf THE TWO-EXPONENTIAL LIOUVILLE THEORY AND}
\centerline{\bf THE UNIQUENESS OF THE THREE-POINT FUNCTION}
\bigskip
\bigskip
\centerline {L.~ O'Raifeartaigh\footnote{${}^1$}{e-mail: lor@stp.dias.ie}, 
J.~M.~ Pawlowski\footnote{${}^2$}{e-mail: jmp@stp.dias.ie}, and V.~V.~ 
Sreedhar\footnote{${}^3$} {e-mail: sreedhar@stp.dias.ie}} 
\centerline {School of Theoretical Physics} 
\centerline {Dublin Institute for Advanced Studies}  
\centerline {10 Burlington Road, Dublin 4, Ireland}
\bigskip
\bigskip
\bigskip
\centerline {\bf Abstract}

It is shown that in the two-exponential version of Liouville theory the
coefficients of the three-point functions of vertex operators can be determined
uniquely using the translational invariance of the path integral measure and 
the self-consistency of the two-point functions. The result agrees with that 
obtained using conformal bootstrap methods. Reflection symmetry and a 
previously conjectured relationship between the dimensional parameters of the 
theory and the overall scale are derived.
\vfil\eject
\noindent {\it Introduction:} Although the quantisation of two-dimensional 
Liouville theory with a potential of the form $V_b=\mu e^{2b\tilde\phi(x)}$, 
where $\tilde\phi(x)$ is a scalar field and $\mu$ and $b$ are constants, has 
been widely investigated [1], it still presents some problems. What is perhaps 
the most disturbing problem is the following: The three point functions of 
vertex operators $\exp[{2\alpha_I\tilde\phi (x_I)}]$ play a central role in the 
theory, in the sense that all $N$-point functions can be obtained from these 
by integration, and have the following form [2, 3]  
$${\cal G}_3 = {\bf C}_3\prod_{I=1}^3\vert{ x_I-x_J\over L}\vert^{\Delta_{(IJ)}
(\alpha)}\quad \hbox{where}\quad {\bf C}_3 = \Bigl(g_b(\xi)
Z_0^b(\xi)\Bigr) \Bigl({K(\xi, \alpha_I )\over k'(-\xi)}\Bigr) \eqno(1)$$ 
Here $L$ is a constant scale, the $\Delta_{(IJ)}$ are known combinations of 
constant conformal weights, $\xi=q-\sum_{I=1}^3\alpha_I$, and $q=b+1/b$, 
while $Z_0^b(\xi)$ is related to the zero-mode integration and can be computed 
explicitly.  $K$ and $k$ are the functions defined in (26). The problem is that
${\bf C}_3$ can actually be computed only at the points $\xi=mb$ for $m\in Z_+$
and the extrapolation from these points to general values of $\xi$ leaves a 
factor $g_b(\xi)$ in (1) undetermined. So far it has not been possible to 
determine this factor from first principles. Instead, what has been done is to 
make the so-called DOZZ Ansatz that 
$${\bf C}_3={K(\xi,\alpha_I)\over k(-\xi)}~~~\Rightarrow ~~~h_b(\xi ) \equiv 
~g_bZ_0^b{k\over k'}~ = 1 \eqno(2)$$ 
and check that the resulting three-point function satisfies some reasonable, 
but extra, conditions such as reflection symmetry and crossing symmetry. The 
latter check is the decisive one, but uses some special {\it four}-point 
functions [4]. 

In a previous paper [5] it was suggested that the gap in the direct derivation 
of $h (\xi)$ could be closed by using a potential of the form $V(\phi)= 
\mu_be^{2b\tilde\phi}+ \mu_ce^{2c\tilde\phi}$, where $bc=1$, -- this being the 
most general potential one can use in the path integral context for a conformal
field theory of a single real scalar field without derivative interactions [6] 
-- rather than the standard form $V_b$. As in the one-exponential theory, the 
${\bf C}_3$ in the two-exponential theory can be calculated only at a discrete,
but much larger, set of points, namely $\xi_{mn} = mb + nc$ and the equation 
corresponding to (1) takes the form
$${\bf C}_3= \Bigl(g(\xi)Z_0(\xi)\Bigr)\Bigl({K(\xi , \alpha_I)\over
k'(-\xi)} \Bigr)\eqno(3)$$ 
where $g(\xi)$ is {\it a priori} undetermined, the functions $K$  and $k$ are
exactly the same as in (1) but $Z_0$ is not. It was shown in [5] that, subject 
to two conditions, the factor $g(\xi)$ could be fixed and thus the DOZZ Ansatz 
(2) could be {\it derived}. The conditions were (a) that the dimensional 
parameters $\mu_b$ and $\mu_c$ were related to the overall scale $L$ of the 
system by an equation of the form $\Omega(\mu_b,\mu_c, L)=1$, where $\Omega$ 
is the function defined in (25), and (b) that $h(\xi)$, defined as 
$h(\xi)=g(\xi)Z_0(\xi)k(-\xi)/k'(-\xi)$, had no singular points. 

In the present paper we refine these results considerably and extend the 
analysis of the two-exponential theory. In particular we show that, subject to 
the mild technical condition given in (37), the DOZZ Ansatz follows from
the translational invariance of the path integral measure and the 
self-consistency of the two-point functions. We also present an extrapolation 
of the fluctuating part of the path integral from the lattice points 
$\xi_{mn}$, for which the $Z_0$ part of the path integral is automatically  
$$Z_0(\xi)={k'(-\xi)\over k(-\xi)} \eqno(4)$$ 
This means that {\it the zero mode integral for the two-exponential theory 
produces exactly the factor $k'/k$ that was postulated in the DOZZ Ansatz}.  
 
As in [5], we use the path-integral formalism; and as the symmetries of the 
path-integral and the associated sum rules are of interest in their own right, 
a secondary purpose of the paper is to present these in a systematic way. The 
most important symmetries are those connected with the translational invariance
of the measure and conformal (Weyl) invariance.  

A by-product is an analysis of the two-point function ${\cal G}_2$. Although 
${\cal G}_2$ cannot be defined directly because of conformal invariance, it can 
be defined both as a limit of ${\cal G}_3$ when one of the $\alpha$'s becomes 
zero, and as a volume integral of ${\cal G}_3$ when the corresponding $\alpha$ 
is $b$ or $c$. The compatibility of the two definitions and the sum rule 
mentioned above lead to a simple linear homogeneous sum rule for the quantity 
$h(\xi)$. Together with the boundary conditions $h(\xi_{mn})= \Omega^{m+n}$, 
obtained by direct computation, this sum rule fixes $h(\xi) = 1$ uniquely. It 
also fixes $\Omega = 1$ which is the relation between the dimensional 
constants $\mu_b$ and $\mu_c$ and the overall scale $L$ that was conjectured 
in an earlier paper. The corresponding sum rule and boundary conditions in the 
one-exponential theory would imply only that $h(\xi)$ is periodic. If 
${\cal G}_2$ is interpreted as an inner-product of states, then $h(\xi)=1$ 
implies that for each conformal weight a zero-norm state decouples, so that 
there is only one physical state. The decoupling is  equivalent to reflection 
invariance, which therefore emerges as an output.  

\noindent {\it The generating functional:} The generating functional of the 
two-exponential Liouville theory is defined as
$$Z[J]= \int [d\tilde\phi]\, e^{-\int d^2x\sqrt{g(x)}~\bigl[{1\over 4\pi}
\tilde\phi \Delta\tilde\phi + {q\over 4\pi}{\cal R}\tilde\phi+ \mu_b e^{2b
\tilde\phi} + \mu_c e^{2c\tilde\phi} - J(x)\tilde\phi(x) \bigr]} \eqno(5)$$  
where $\Delta$ is the Laplace-Beltrami operator, ${\cal R}$ is the Ricci 
scalar, the coupling constants $\mu_b, \mu_c$ have dimensions of mass squared, 
and $b$ and $c$ are dimensionless constants. For the $N$-point functions of 
vertex operators we have $e^{\int d^2x\sqrt g J\tilde\phi} = \prod_{I=1}^N
\nu_I\sqrt g e^{2 \alpha_I\tilde\phi}$, but we need not yet specialise to this
case. The $\nu_I$ are constants which, like the $\mu_b$ and $\mu_c$, have to 
be renormalised because of the short distance singularities of the Green's 
function defined by the equation $\Delta G(x,y) = {\pi\over\sqrt g}
\delta^2(x-y)$. As explained in detail in [5], the Green's function for 
the non-coincident and coincident arguments is defined as:
$$ G(x, y ) = -{1\over 2}{\hbox{ln}}{\mid x - y \mid\over L} + {\cal O}(V^{-1})
~~{\hbox{and}}~~~G(x,x+dx)\equiv -{1\over 2}\ln [{ds\over L} ] + {1\over 4}\ln 
\sqrt{g(x)} \eqno(6)$$ 
respectively, where $ds$ is the infinitesimal geodesic separation, $L$ 
sets the overall scale, and $V$ is the volume of the space. 

As the Green's function $G(x, x )$ occurs only in the form $-\alpha^2 G(x, x )$ 
at each vertex, its infinite part may be absorbed by renormalising $\mu_b$, 
$\mu_c$ and $\nu_I$. Let us define the renormalisations of $\mu_b$, $\mu_c$ 
and $\nu_I$ to be
$$\mu_b \rightarrow \mu_b(\Lambda {ds\over L})^{2(b^2 -q b)},~~~\mu_c
\rightarrow\mu_c(\Lambda {ds\over L})^{2(c^2 - q c)},~~~\nu_I\rightarrow 
\nu_I (\Lambda {ds\over L})^{2(\alpha_I^2  -q\alpha_I)}\eqno(7)$$
where $\Lambda$ is the dimensionless ultra-violet renormalisation scale.
Anticipating the result that $b + c = q,~~bc = 1$ from Weyl invariance, it 
is easy to see that $\mu_b$ and $\mu_c$ have the naive scaling dimensions with 
respect to $\Lambda$.\footnote{$~^1$}{There is arbitrariness in the definition 
of renormalisation associated with the translational invariance of the path 
integral measure. However, this arbitrariness does not affect equation (8).}  
Clearly the $b^2$, $c^2$, $\alpha^2$ powers of $ds$ in the above equation 
cancel the infinities coming from the Green's functions at coincident points. 
Counting the powers of the remaining factors of $ds$ and $\Lambda$, we find, 
after a little algebra, that the renormalised functional integral is 
proportional to 
$$[{ds\over L}]^{-2q^2}\prod_{I=1}^N\Lambda^{2(\alpha_I^2 -q\alpha_I)}\eqno(8)$$
Since the factor $[ds/L]^{-q^2}$ may be absorbed in the Polyakov conformal 
anomaly term which is proportional to $q^2$, we see that the $\Lambda$ 
dependence of the renormalised functional integral cancels except for the 
contribution coming from the external sources. 

The conformal weights $\Delta_\alpha$ of the fields exp [$2\alpha_I\tilde\phi 
(x_I)$] may be read off directly by varying the path integral with respect 
$\sqrt{g (x_I)}$. Taking the $\sqrt g$ dependence of $\alpha_I^2G(x_I, x_I)$ 
and the cross-term  $q\alpha_I\int d^2x\int d^2y\sqrt {g(x)}{\cal R}(x)G(x, y)
\delta (y - x_I )$ in the Gaussian integration into account, this yields 
$\Delta_\alpha = \alpha ( q- \alpha )$. From (7) it is clear that the scaling 
dimensions of $\nu_I$ are minus the conformal weight of corresponding field. 
Thus fields of the same conformal weight are renormalised in the same way. 

\noindent {\it A sum rule for the generating functional:} If we now specialise 
to the case where the external current is of the form $\prod_{I=1}^N\nu_I 
e^{\int d^2x\sqrt g J\tilde\phi} = \prod_{I=1}^N\nu_I\sqrt g e^{2\alpha_I\tilde
\phi}$, the renormalised functional integral is defined by 
$$Z[J]= \int [d\tilde\phi ]~e^{- \int d^2x\sqrt{g(x)}~ \left[{1\over 4\pi}
\tilde\phi\Delta\tilde\phi + {q\over 4\pi}{\cal R}\tilde\phi+ \sqrt{g}^{b^2}\,
\mu_b e^{2b\tilde\phi}+\sqrt{g}^{c^2}\,\mu_c e^{2c\tilde\phi}\right]}
\prod_{I=1}^N\sqrt {g}^{\alpha_I^2} \nu_Ie^{2\alpha_I\tilde\phi(x_I)}\eqno(9)$$ 
where we are using the renormalisation prescription in (7) and we have 
absorbed, for convenience, the ln$\sqrt g$ part of $G(x,x)$ by letting 
$e^{2\alpha\phi}\rightarrow {\sqrt g}^{\alpha^2}e^{2 \alpha \phi}$. It is then 
understood that $G_R(x, x) = 0$. 

The translation invariance of the path integral measure can now be formulated 
as follows:    
$$ \int [d\tilde\phi] ~{\delta\over \delta \tilde\phi(x)}\left[e^{-S[\tilde
\phi]+\int d^2 x\sqrt{g} J(x)\tilde\phi(x)}\right]=0\eqno(10)$$
where ${\delta\over \delta \tilde\phi(x)}$ is the generator of translations on 
the space of fields. From (10) we derive the quantum equations of motion for 
$Z[J]$: 
$${1\over 4\pi}\Delta{\delta Z[J]\over\delta J(x)}=-\Bigl(b\mu_b{\sqrt g}^{b^2}
Z[J_{b,x}] +c\mu_c {\sqrt g}^{c^2}Z[J_{c,x}]\Bigr)+{1\over 2}\bigl(J(x)-{q\over
4\pi}{\cal R}(x) \bigr)Z[J]\eqno(11)$$
where 
$$J_{b,x}(y)= J(y)+2b {1\over \sqrt{g(y)}}\delta^2(y-x), \quad J_{c,x}(y)= 
J(y)+2c {1\over \sqrt{g(y)}}\delta^2(y-x)\eqno(12)$$ 
Carrying out the same procedure for the constant, zero-mode, part $\phi_0$ of 
$\tilde\phi$, we obtain the integrated form of (11), namely  
$$\int d^2 x\sqrt{g} \Bigl(b\, \mu_b{\sqrt g}^{b^2} Z[J_{b,x}]+c\, \mu_c {\sqrt
g}^{c^2}Z[J_{c,x}]\Bigr)={1\over 2}\left(J_0-q\chi\right)Z[J] \eqno(13)$$
where 
$$J_0=\int d^2 x \sqrt g J(x)  \quad {\hbox{and}} 
 ~~~\chi={1\over 4\pi}\int d^2 x\sqrt{g} {\cal R}(x)\eqno(14) $$
$\chi$ being the Euler number of the underlying manifold. Eq. (13) embodies 
the sum rule for the generating functional.  

\noindent {\it Weyl transformations:} A local Weyl transformation can be 
performed by varying the generating functional with respect to $\sqrt {g (x)}$ 
where $x \neq x_I$, the external points. For $Z[J]$ in (9), we find 
$${\delta Z\over \delta\sqrt g} = (1+b^2){\sqrt g}^{b^2} \mu_b\, Z[J_{b,x}]+
(1+c^2){\sqrt g}^{c^2} \mu_c\, Z[J_{c,x}]+ {q\over 4\pi}\Delta{\delta Z[J]\over
\delta J(x)} \eqno(15)$$ 
The third term may be eliminated using (11) to get  
$$\Bigl(1+b^2-qb\Bigr){\sqrt g}^{b^2}\mu_b Z[J_{b, x}]+ \Bigl(1+c^2-qc\Bigr)
{\sqrt g}^{c^2}\mu_c Z[J_{c, x}] - \left({q\over 4\pi}{\cal R}(x) - J(x)\right)
{q\over 2}Z(J)\eqno(16)$$ 
The Weyl condition is that the variation ${\delta Z\over\delta \sqrt g}$ should
be proportional to the external current namely ${q\over 4\pi} {\cal R} - J$. 
Since this has to be  valid for all currents $J$, the condition for Weyl 
invariance is that the first two terms must vanish and  we have 
$$q=(b+c)~~~{\hbox{and}}~~~ b c=1, \eqno(17)$$ 
The above approach may be contrasted with the one in [5] where (17) was derived 
only after the path integral was evaluated. 

\noindent {\it The $N$-point functions:} For the computation of general 
$N$-point functions of vertex operators, we let the underlying manifold be a 
two dimensional sphere. In that case, $\chi = 2$ and there is only one 
zero-mode for $\tilde\phi$, namely the constant $\phi_0$. As explained in 
detail in [5], the expression for the $N$-point function may be simplified by 
using a Sommerfeld-Watson transform [7] for the exponential of an integrated 
vertex operator. With $\tilde\phi=\phi_0+\phi$ the resulting expression for the 
$N$-point function takes the form   
$${\cal G}_N = \int d\phi_0\int {du\,dv\,\over\Gamma(1+iu)\Gamma(1+iv)}
{e^{2(i(bu+cv)-\xi_N)\phi_0}\over\hbox{sinh}\pi u~\hbox{sinh}\pi v}
\times {\bf C}_N(iu,iv)\eqno(18)$$
where 
$${\bf C}_N(iu,iv) = \int d\phi\,U_b^{iu}\, U_c^{iv}\, e^{-\int d^2x\sqrt{g}
\bigl[{1\over 4\pi}\phi \Delta\phi + {q\over 4\pi}R\phi\bigr]}{\bf \Pi}_N ~~~
{\hbox{with}}~~~\xi_N = q - \sum_{I =1}^N\alpha_I\eqno(19)$$ 
and 
$$ {\bf \Pi}_N = \prod_{I=1}^N {\sqrt g}^{\alpha_I^2}\psi_Ie^{2\alpha_I
\tilde \phi(x_I)},~~~~U_b (\phi ) = \mu_b\int d^2x (\sqrt{g})^{qb}e^{2b\phi}
\eqno(20)$$ 
and similarly for $b\leftrightarrow c$. The integral ${\bf C}_N(iu,iv)$ is 
a Gaussian path integral for the fluctuations $\phi$ which, for $iu=m$ and 
$iv=n$, $m$ and $n$ being positive integers, can be done in a straightforward 
manner and produces an ordinary multiple integral. 

In terms of the $N$-point functions ${\cal G}_N(x_I,\alpha_I)$, the sum rule 
(13) takes the form 
$$\int d^2 x \sqrt g \left[b\mu_b {\cal G}_{N+1}(x_I,x,\alpha_I,b)+ 
c\mu_c{\cal G}_{N+1} (x_I,x,\alpha_I,c)\right]=-\xi_N {\cal G}_N(x_I,\alpha_I)
\eqno(21)$$
and thus relates the (integrated) $N+1$-point function to the $N$-point 
function. Note that the above equation requires that $\xi_N\neq 0$ because 
${\cal G}_N\rightarrow \infty ~{\hbox{as}}~ \xi_N\rightarrow 0$, making the 
right hand side indefinite.      

\noindent{\it The Three-point function:} As is well-known, the three-point 
function is the lowest $N$-point function for which conformal invariance does 
not require the extraction of an infinite group volume factor. If we choose 
$\xi$ to be pure imaginary and integrate over the zero-mode $\phi_0$ in (18) 
we obtain a delta function $\delta(\xi-bu-cv)$, in which case the coefficient 
${\bf C}_3$ may be written as 
$${\bf C}_3(\xi, i(u+v)) = \int d\phi e^{-{1\over 4\pi} \int d^2x\sqrt{g}
[\phi\Delta\phi+ qR\phi]}U_b^{iu}U_c^{iv}{\bf\Pi_3}\eqno(22)$$
Apart from the spectator variables $\alpha_I -\alpha_J$, we see that, due to 
the delta-function, ${\bf C}_3$ is a function of only two variables, chosen as 
$\xi$ and $u+v$ for convenience. This is the great advantage of using the 
Sommerfeld-Watson transform. We then have in the infinite volume limit, 
$${\cal G}_3 = \int {dudv\over\Gamma(1+iu)\Gamma(1+ iv)}{\delta 
(\xi - ibu - icv)\over\hbox{sinh}\pi u~\hbox{sinh}\pi v}{\bf C}_3(\xi, i(u+v))
\eqno(23)$$
The problem is that ${\bf C}_3$ can only be computed at the points 
$u,v= -im, -in$ for $m,n\in Z_+$, where, as shown in [5], it is given by,  
for $\xi_{mn}\equiv bm+cn$, 
$${\bf C}_3(\xi_{mn},m+n)= (-1)^{m + n}m!n!\lambda^{\xi_{mn}}\Omega^{m+n}
\Bigl({K(\xi_{mn},\alpha_I)\over k'(-\xi_{mn})}\Bigr)
\prod_{I=1}^3\mid{x_{IJ}\over L}\mid^{\Delta_{IJ}(\alpha )}
\eqno(24)$$
where 
$$\lambda = \Bigl({\mu_b\Phi_b\over \mu_c\Phi_c} \Bigr)^{1\over b-c},~~~
\Omega^{c - b}={(\mu_bL^2\Phi_b)^c\over (\mu_cL^2\Phi_c)^b},~~~
\Phi_b=\pi\gamma(b^2)(b^2)^{2-qb}, ~~~\gamma (x)={\Gamma(x)\over \Gamma(1-x)}
\eqno(25)$$
The functions $k(\xi )$ and $K(\xi, \alpha_I)$ are defined by the following 
equations:
$$\ln k (\xi ) = \int_0^\infty {dt\over t}\Bigl({({q\over 2} - x)^2 e^{-2t}
-{{\hbox{sinh}}^2({q\over 2} - x)t\over{\hbox {sinh}}bt~{\hbox{sinh}}ct}}\Bigr)
~{\hbox {and}}~K(\xi,\alpha_I) = k'(0)\prod_{I=1}^3{k(2\alpha_I)\over 
k(\xi + 2\alpha_I)} \eqno(26)$$
To extrapolate ${\bf C}_3$ to other values of $\xi$ we note that although the 
arguments of $k$'s are constrained by the relation $\sum_I \alpha_I = q -
\xi_{mn}$, they range over the whole real axis for fixed $\xi_{mn}$. Hence the 
only reasonable extrapolation is the obvious one, $k(\xi_{mn},\alpha_I)
\rightarrow k(\xi, \alpha_I)$, in which case $K(\xi_{mn},\alpha_I)\rightarrow 
K(\xi,\alpha_I)$. Unfortunately, this argument is not valid for the rest of 
${\bf C}_3(\xi_{mn}, m+n)$, which depends only on $\xi_{mn}$. Thus the most 
general extrapolation of (24) is 
$${\bf C}_3 (iu,  iv, \xi ) = {\hbox{cosh}}\pi u~{\hbox{cosh}}\pi v\Gamma(1+iu)
\Gamma(1+iv)\lambda^\xi\Omega^{i(u+v)}{K(\xi,\alpha_I)\over k'(-\xi)}
f(\xi,u+v) \mid {x_{IJ}\over L}\mid^{\Delta_{(IJ)}}\eqno(27)$$
where the cosh terms take care of the $(-1)^{m + n}$ terms and $f(\xi,u+v)$ is 
an arbitary function with $f(\xi = mb+ nc , m+n) = 1$. Then ${\cal G}_3$ 
becomes 
$${\cal G}_3= \lambda^\xi h(\xi) {K(\xi,\alpha_I)\over k(-\xi)}\prod_{I=1}^3
\mid {x_{IJ}\over L}\mid^{\Delta_{(IJ)}(\alpha_I )} \eqno(28)$$
where 
$$h(\xi)={k(-\xi)\over k'(-\xi)}Z_0(\xi) \quad \hbox{and} \quad Z_0(\xi)=\int 
{dudv~\delta(\xi-ibu-icv)\over\hbox{tanh}\pi u~\hbox{tanh}\pi v}\Omega^{i(u+v)}
f[\xi, i(u+v)] \eqno(29)$$ 
Since the function $f$ is unknown, we cannot proceed from (29). Hence we 
take an alternative route using the two-point function. 

\noindent {\it Uniqueness (An Application of the Sum Rule):} For two and three 
point functions the sum rule (21) in the infinite volume limit is 
$$b\mu_b\int d^2 x {\cal G}_{3}(x_I,x,\alpha_I,b)+ c\mu_c\int d^2x {\cal G}_{3}
(x_I,x,\alpha_I,c)=-\xi_2 {\cal G}_2(x_I,\alpha_I),\qquad \xi_2\not=0\eqno(30)$$
This is not useful unless we have an alternative definition for the two-point 
function. Such a definition may be obtained by regarding it as 
twice\footnote{$~^2$}{Actually any constant independent of $\xi$ is allowed
{\it a priori}, but the requirement that the sum rule be satisfied at the 
points $\xi = mb +nc$ fixes the constant to be two.} the limit of the 
three-point function as $\alpha_3\rightarrow 0$ and $x_3\rightarrow\infty$. It 
is easy to see that the limit is non-zero only if $\Delta_1 = \Delta_2$ {\it 
i.e} if $\alpha_1 = \alpha_2$ or $\alpha_1 = q - \alpha_2$. Since for 
scattering states $\alpha_I= {q\over 2} + i\beta_I$, the quantities 
$\alpha_1- \alpha_2$ and $\xi_2=q- \alpha_1-\alpha_2$ are pure imaginary, and 
we obtain 
$${\cal G}_2(\alpha_1,\alpha_2;x_{12})= 4\pi\lambda^{\xi_2} \mid{x_{12}\over L}
\mid^{-(\Delta_1+\Delta_2)}h(\xi_2) R(\xi_2) [\delta(\beta_1-\beta_2) + 
\delta(\beta_1+\beta_2)]\eqno(31) $$
where $R(\xi)= k(\xi)/ k(-\xi)$. 
 
To compute the left hand side of (30) we note that the first integral is 
$$\lambda^{\xi_2-b}h(\xi_2 -b){K(\alpha_1,\alpha_2,b)\over k(-\xi_2 +b)}
\vert{x_{12}\over L}\vert^{2(\Delta_b-\Delta_1-\Delta_2)}\times\zeta\eqno(32)$$
where 
$$\zeta =\int d^2x_3 \vert {x_{31}\over L} \vert^{-2(\Delta_b+D)}\vert {x_{23}
\over L} \vert^{-2(\Delta_b-D)}= 2\pi^2 \mid {x_{12} \over L} \mid^{-2\Delta_b}
\delta(D)\eqno(33) $$
and $\Delta_b = 1$ and $D = \Delta_1 - \Delta_2=\beta_1^2-\beta_2^2$. The delta
functions in these equations and the condition $\xi_2\not=0$ in (30) mean that
we only need the coefficient for $\alpha_1 = \alpha_2\equiv \alpha$, which is  
easily computed to be
$${K(\alpha,\alpha,b)\over k(-\xi_2 +b)}= -\xi_2^2 k'(0)R(\xi_2){k(2b)\over k^2
(b)} = \xi_2^2 R(\xi_2 ){\Phi_b\over \pi b},~~~~\xi_2 = q - 2\alpha\eqno(34)$$
Inserting these formulae into the sum rule (30) and using the identity 
$\lambda^{b}=\mu_b\Phi_b\Omega$, and similarly for $c$, we get the sum rule 
$$ h(\xi_2 + b) +  h(\xi_2 + c) = 2\Omega h(\xi_2 )~~~{\hbox{with}}~~~
h(mb+nc)= \Omega^{m+n} \eqno(35)$$
for the unknown function $h(\xi_2)$. The second equation is  obtained by 
explicit computation from (24). The question is whether the equations in (35) 
determine $h(\xi_2)$ uniquely. We show that they do subject to a mild
technical assumption to be introduced presently. First let us restrict
ourselves to the case when $b$ is fractional i.e. $b=r/s$ where $r>s$ are 
positive integers with no common factor and lowest common multiple $rs$. Next 
we rewrite the sum rule (35) in terms of new variables defined as follows: 
$$ y \equiv e^\xi~~~ \Rightarrow ~~~h(e^{-{r\over s}}y) + h(e^{-{s\over r}}y) 
= 2\Omega h(y) \eqno(36)$$ 
We now make the assumption\footnote {$^3$}{It is probable that this could be  
actually proved directly from the sum rule (35) which rules out a whole 
class of functions.} 
$$\lim_{y\rightarrow 0} {h\bigl(e^{-({r\over s})^{\pm 1}} y \bigr)\over h(y )} 
~\equiv~ C_\pm < \infty \eqno(37)$$
Considering the restriction 
$$y_t = e^{t\over rs}~~~ \Rightarrow ~~~h(e^{t - r^2\over rs}) + 
h(e^{t - s^2\over rs}) = 2\Omega h(e^{t\over rs}),~~~{\hbox{where}}~~~t\in Z_+
\eqno(38)$$
we see that the symmetry group we need to implement is the dilatation group.
Thus we may expand a solution of (38) as follows:
$$ h(y_t) = \sum_{p= 0}^N C_p y_t^{\sigma_p}~~~
{\hbox {where}}~~~ e^{-r^2\omega\sigma_p } + e^{-s^2\omega\sigma_p} - 2\Omega =
0 ~~~{\hbox{and}}~~~\omega = {1\over rs} \eqno(39)$$
$N ~(0\leq N\leq r^2)$ being a finite number. This follows because the sum rule 
restricts the number of $C_p$s to be the dimension of the solution space of the 
polynomial equation in (39). This is of course true only on the first sheet of 
the covering of the $y$ variable. In the general case we would also have a sum 
over the infinite number of coverings. 
$$h(y) = \sum_{p = 0}^N\sum_{n= -\infty}^\infty C_{p_n}y^{\sigma_{p_n}}
~~~{\hbox {where}}~~~\sigma_{p_n}= \sigma + 2\pi inrs\eqno(40)$$
We now use the second equation in (35) to write, for the special values  
$\xi = mb + nc \Rightarrow t = mr^2 + ns^2$, 
$$\sum_{p= 0}^N C_p y_t^{\sigma_p} = \Omega^{m+ n} \Rightarrow 
C_N = 1,~~\sigma_N = 0,~~ C_{p\neq N} = 0\Rightarrow \Omega = 1 \eqno(41)$$
The latter follows as the only consistent solution since the finite sum on the 
left hand side cannot in general match the right hand side which can be made 
arbitrarily large by letting $m$ and $n$ tend to infinity. Note that the 
covering does not play a role in the discussion of (41). It then follows that 
$$h ({t\over rs}) = 1~~~ \forall~~ t\in Z_+\eqno(42)$$ 
Thus for $b = r/s$ the function $h(\xi )$ is unity when restricted to the 
subset of points $\xi = t/rs$. 

This is as far as we can go for a fixed $b=r/s$. However we now invoke the 
fact that (29) implies that $h(\xi)$ is continuous in $b$ and $\xi$. We then 
define $\tilde b=\tilde r/\tilde s$ where $\tilde r=lr+1$ and $\tilde s=ls$. It
follows that $\tilde r$ and $\tilde s$ have no common factor, $\tilde b-b=1/ls$
and, by applying the above result to the tilde-variables, $h(t/l^2sr(1+1/lr)
=1$. But this means that in any $1/l$ neighbourhood of $b$ there is a $\tilde 
b$ for which $h(\xi)=1$ at points which are separated by distances of order 
$1/l^2$. As these distances tend to zero as $l$ tends to infinity, we see that 
this is compatible with the continuity of $h(\xi)$ in $b$ and $\xi$ only if 
$h(\xi)=1$ for all $\xi$.   

\noindent {\it  Reflection symmetry:} Once $h(\xi)=1$ it follows that the 
denominator in the three-point function is invariant under the reflection 
$\alpha_I\rightarrow q-\alpha_I$ for each $I$ and thus the three-point 
function is covariant with respect to reflection symmetry in the sense that 
$${\cal G}_3 (q - \alpha_1,\alpha_2, \alpha_3) = R(q - 2\alpha_1){\cal G}_3
(\alpha_1, \alpha_2, \alpha_3) \eqno(43) $$
where the prefactor depends only on the reflected parameter $\alpha_1$. Thus 
in the two-exponential theory, reflection covariance is an output rather than 
an input.  

It is interesting to note how this reflection covariance expresses itself in 
terms of the two-point function defined. If we interpret the two-point function
normalised by the volume factor $4\pi$, as the inner product of primary states  
$$ <\alpha_1,\alpha_2> \equiv \lim_{x\to 0}{1\over 4\pi }\mid x\mid^{2
\Delta_\alpha} {\cal G}_2(q- \alpha_1,\alpha_2;x) \eqno(44)$$
we have  
$$\pmatrix{ <\alpha,\alpha >& <\alpha, q-\alpha >\cr <q-\alpha,\alpha > & <q - 
\alpha,q-\alpha >} =\pmatrix{1&R^{-1}(q - 2\alpha)\cr R(q - 2\alpha )&1}
\delta(0) \eqno(45)$$
It is clear that the matrix in (45) is hermitian and has zero determinant. 
Hence one linear combination of the states, namely $\vert\alpha > - R(q - 2
\alpha )\vert  q-  \alpha > $, has zero norm and decouples. Thus effectively, 
$$\vert \alpha >~~ = ~~R(q - 2\alpha )\vert q - \alpha >\eqno(46) $$
which means that there is actually only one physical state for each conformal 
weight $\Delta_\alpha$. This may seem surprising but from (43) it is seen to 
be a manifestation of the reflection covariance. 

\noindent {\it Comparison of the one and two-exponential path integrals:} In 
order to compare the one and two exponential theories, we begin by recalling
that ${\bf C}_3$ in both the theories is defined in terms of the correlation 
functions of vertex operators in a free field theory. In the two-exponential
theory, the relevant integral is given by 
$$\int d\phi e^{-S(\phi )}U_b^{iu}U_c^{iv}{\bf \Pi}_3 = 
\Gamma (1 + iu) \Gamma (1 +iv){K(-\xi, \alpha_I )\over k'(-\xi )}
f[\xi, i(u + v)]\eqno(47)$$ 
where $S(\phi )$ is the Action for the free theory. The corresponding equation,
in the one-exponential theory, is obtained by letting $v\rightarrow 0$ and 
takes the form 
$$\int d\phi e^{-S(\phi )}U_b^{iu}{\bf \Pi}_3 = \Gamma (1 + iu){K(-\xi, 
\alpha_I )\over k'(-\xi )}\times f[\xi, iu]\eqno(48)$$
The integral corresponding to the zero-mode integral in (29) in the 
one-exponential theory for an {\it a priori} arbitrary $f[\xi , iu]$ can be 
performed to yield $f[\xi , c\xi ]/{\hbox{tanh}}\pi c\xi $. If we {\it assume}
that the one-exponential theory leads to the DOZZ Ansatz, then the function 
$f[\xi , iu]$ is determined {\it uniquely} to be $b{{\hbox{tan}}(\pi u) 
k'(-ibu )/k(-ibu)}$ for $\xi = ibu$. We may now ask, what choice, if any,
for the function $f[\xi, i(u+ v)]$ in the two-exponential theory will 
produce the DOZZ Ansatz. It is easy to see that if we choose 
$$ f[\xi , i(u + v)] = {k'(-\xi )\over k(-\xi )}[b{\hbox {tanh}}\pi u + 
c{\hbox {tanh}}\pi v]\eqno(49)$$ 
the $Z_0$ integral in (29) produces $k'/k$ in accordance with the DOZZ Ansatz.
This choice has the virtue that it reduces to the one-exponential result 
in the limit $v\rightarrow 0$. However, one can easily convince oneself that 
there exist other choices for the function $f[\xi , i(u + v)]$. If we choose, 
for example, $f[\xi, i(u+v)] = {\hbox{sgn}}[i(u + v)]$ (the sgn-factor is 
necessary to preserve the symmetry of the path integral under a change of sign 
of $b,c$ and the $\alpha$'s) and integrate over $u$ and $v$ first, we obtain,
$$Z_0(\xi)=\int_0^\infty  dt {\hbox{sinh}((q-2\xi)t)\over\hbox{sinh}(bt)
\hbox{sinh}(ct)},\hskip 2truecm t\equiv \vert\phi_0\vert \eqno(50)$$
Up to a regulating term,\footnote {$~^4$}{The regulating term can be obtained 
naturally by modifying the path-integral measure $d\phi_0$ to $\lim_{\epsilon
\rightarrow 0}d\phi_0\bigl[1-\hbox{exp}(-\vert \phi_0\vert/ \epsilon)\bigr]$.} 
this will be recognized, from the definition of $k(\xi)$ in (26), as $k'(-\xi)/
k(-\xi)$. Since the choice for the extrapolation function is not unique we 
conclude that any result that is based on an extrapolation is not conclusive.
It is therefore desirable to have a direct proof of the DOZZ result for the 
three-point function without any reference to the extrapolation. This is 
exactly what was achieved in this paper by showing that the sum rule in the 
two-exponential theory has a unique (constant) solution. In contrast, in the 
one-exponential theory, if the specific extrapolation leading to the DOZZ 
Ansatz is not made, the final result can only be obtained up to a periodic 
function of $\xi$. 
\bigskip
\noindent {\it Acknowledgements:} We thank I. Sachs, J. Teschner and 
P. Watts for useful discussions. 
\vfil\eject
\centerline {\bf References}

\item{1. } H. Poincare, J. Math. Pure App. {\bf 5} {\it se 4} (1898) 157;
N. Seiberg, Notes on Quantum Liouville Theory and Quantum Gravity, in 
"Random Surfaces and Quantum Gravity", ed. O. Alvarez, E. Marinari, 
and P. Windey, Plenum Press, 1990; A. Polyakov, Phys. Lett. {\bf B103} (1981)
207; T. Curtright and C. Thorn, Phys. Rev. Lett. {\bf 48} (1982) 1309; 
E. Braaten, T. Curtright and C. Thorn, Phys. Lett {\bf 118} (1982) 115; Ann. 
Phys. {\bf 147} (1983) 365; J.-L. Gervais and A. Neveu, Nuc. Phys. {\bf B238} 
(1984) 125; {\bf B238} (1984) 396; {\bf 257}[FS14] (1985) 59; E. D'Hoker and 
R. Jackiw, Phys. Rev. {\bf D26} (1982) 3517.
\item{2. } H. Dorn and H.-J. Otto, Phys. Lett. {\bf B291} (1992) 39;
Nuc. Phys. {\bf B429} (1994) 375.
\item{3. } A. and Al. Zamolodchikov, Nuc. Phys. {\bf B 477} (1996) 577; 
Low Dimensional Applications of Quantum Field Theory,
ed. L. Baulieu, V. Kazakov, M. Picco, and P. Windey, NATO ASI Series B,
Physics Volume 362, Plenum Press 1997.     
\item{4. } J. Teschner, Phys. Lett. {\bf B363} (1995) 65.
\item{5. } L. O'Raifeartaigh, J. M. Pawlowski, and V. V. Sreedhar, 
Annals of Physics {\bf 277} (1999), 117. 
\item{6. } {\it See also} Vl. S. Dotsenko, Mod. Phys. Lett. {\bf A6} 
(1991), 3601; E. D'Hoker, Mod. Phys. Lett {\bf A6} (1991), 745; P. Mansfield, 
Phys. Lett. {\bf B242} (1990), 242.
\item{7.} G. N. Watson, Proc. Soc. London {\bf 95} (1918), 83; 
A. Sommerfeld, "Partial Differential Equations of Physics", Academic Press, 
New York 1949.  
\item{8. } Vl. Dotsenko and V. Fateev, Nuc. Phys. {\bf B251} (1985) 691.
\vfil\eject\end